\documentclass[usenatbib]{mn2e}
\usepackage[dvips]{color}
\usepackage{graphicx}
\usepackage{lscape}

\def\*{$^{*}$}

\def\deg{$^\circ$}

\def\arcsec{$^{\prime\prime\,}$}
\def\arcmin{$^{\prime\,}$}

\def\flux{ergs cm$^{-2}$ s$^{-1}$}

\title[Deep Hard X-ray Survey of LMC]
{Deep Hard X-ray Survey of the Large Magellanic Cloud}
\author[Grebenev, Lutovinov, Tsygankov,
  Mereminskiy]{S.\,A.\,Grebenev$^{1}$\thanks{E-mail:
grebenev@iki.rssi.ru}, A.\,A.\,Lutovinov$^{1}$, S.\,S.\,Tsygankov$^{2,3,1}$, I.\,A.\,Mereminskiy$^{1}$\\
$^{1}$Space Research Institute, Russian Academy of Sciences,
  Profsoyuznaya 84/32, Moscow  117997, Russia\\
$^{2}$Finnish Centre for Astronomy with ESO (FINCA), University of Turku, V\"ais\"al\"antie 20, FI-21500 Piikki\"o, Finland \\
$^{3}$Astronomy Division, Department of Physics, FI-90014 University of Oulu, Finland}

\begin{document}

\date{Accepted .... Received .....}

\pagerange{\pageref{firstpage}--\pageref{lastpage}} \pubyear{2012}

\maketitle

\label{firstpage}

\begin{abstract}
%\textcolor{green}
{Results of the deep survey of the Large Magellanic Cloud (LMC),
  performed with the \emph{INTEGRAL\/} observatory, are
  presented. The long exposure ($\sim7$ Ms) allowed us to
  detect twenty one sources in this sky region: ten belonging to
  the LMC itself (7 HMXBs, 2 PSRs, 1 LMXB), six of extragalactic
  origin and three belonging to other galaxies from the Local Group
  --- the Milky Way (2 sources) and Small Magellanic Cloud (1
  source). Four new hard X-ray sources of these 21 ones were
  discovered during the survey in addition to IGR\,J05414-6858
  reported earlier; two of them were identified with
  extragalactic objects. We report also for the first time the
  detection of a hard X-ray emission from the Crab-like pulsar
  PSR\,J0537-6910 and identification of the hard X-ray source
  IGR\,J05305-6559 with the high-mass X-ray binary
  EXO\,053109-6609.  }

\end{abstract}


\section{Introduction}
All-sky surveys recently carried out in hard ($>15$ keV) X-rays
by the \emph{INTEGRAL\/} and \emph{Swift\/} observatories have led to
the discovery of hundreds new sources that enlarged the total
number of hard X-ray objects known on the sky by several times
\citep[e.g.][]{kri2010a,bird2010,kri2012,baum2010,cus2010}. These
observatories working quasi-simultaneously
(\emph{INTEGRAL\/} --- since the end of 2002, \emph{Swift} ---
since the end of 2004) are well supplementing each other:
\emph{Swift} has more uniform coverage of the sky, but
\emph{INTEGRAL\/} has an advantage in observations of the most
crowded fields similar to the Galactic Centre and Galactic
plane. During about $10$ years of operation \emph{INTEGRAL\/} also
deeply observed a number of selected sky fields such as the ones around
3C273, M82, Vela and few more.

The Large Magellanic Cloud has been observed with \emph{INTEGRAL\/} many
times in 2003--2004 and 2010--2012. The primary goal of these
observations was searching for lines of the direct-escape
emission from the radioactive decay of $^{44}$Ti in the remnant
of Supernova 1987A \citep{greb2012}. The achieved long exposure
allowed us to study for the first time this nearby galaxy with
an unprecedent sensitivity in hard X-rays. Note that this
exposure was accumulated mainly during last two years thus the
data of observations used in this paper were not taken into
account in the last revisions of the \emph{INTEGRAL/IBIS/ISGRI}
all-sky survey \citep[e.g.,][]{kri2010a,bird2010}.

In this paper we present sky images of the LMC in hard (20--60
keV) and standard (3--20 keV) X-ray energy bands and the
catalogue of all detected sources, including four newly
discovered ones. Also, we present broadband (3--100 keV) spectra
of eight bright X-ray binaries located in this region (for some of
them such spectra are reported here for the first time). The
statistical study of the population of high-mass X-ray binaries
(HMXBs) in the LMC and the active galactic nuclei (AGNs)
detected in this direction is published elsewhere \citep{lut2012}.

\section{Observations and data analysis}

In this work we used the data obtained with the \emph{JEM-X\/}
\citep{lund03} and \emph{IBIS\/ISGRI} \citep{uber03,lebr03} telescopes
aboard the
\emph{INTEGRAL\/} observatory \citep{win03}. The \emph{IBIS\/}
telescope has a relatively wide field of view ($\sim 29^\circ
\times 29^\circ$ at zero response, $9^\circ \times 9^\circ$
fully coded field) and moderate (12\arcmin\ FWHM) angular
resolution, that in combination with the high sensitivity in the
20--60 keV band makes it to be the best instrument for deep
surveys in hard X-rays. The \emph{JEM-X\/} telescope has smaller
field of view (13\deg\ in diameter at zero response), but
slightly better angular resolution ($\sim3\farcm5$). Due to
these reasons it was used as the secondary instrument for the
survey in order to extend it below 20 keV.

The total exposure for \emph{INTEGRAL\/} observations of the LMC
field reaches currently $\sim7$ Ms (the dead time-corrected
exposure of \emph{IBIS/ISGRI\/} for the central part of the
field is $\simeq4.8$ Ms, for \emph{JEM-X\/} the effective
exposure is smaller $\simeq1.8$ Ms due to its smaller field of
view). Reduction of the \emph{IBIS/ISGRI\/} data was carried out
using the methods described by \cite{kri2010b}, reduction of the
\mbox{\emph{JEM-X\/}} data --- with the standard \emph{OSA}
package version 9.0\footnote{http://isdc.unige.ch}. Spectra of the detected sources were reconstructed by direct
measurements of their fluxes in the images extracted in the narrow
energy bands.

\begin{figure}
\includegraphics[width=0.92\columnwidth,bb=55 280 550 710,clip]{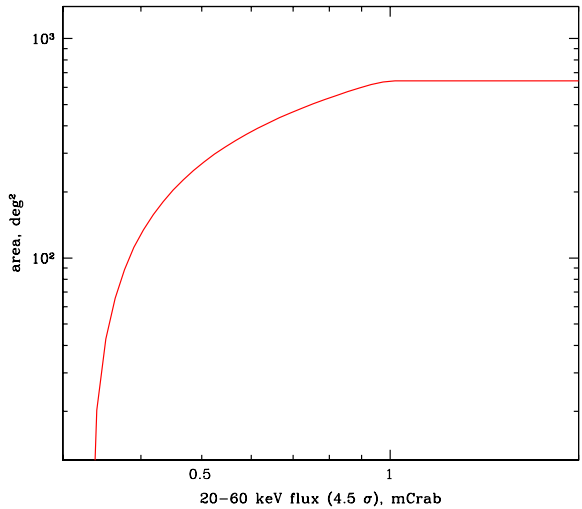}

\caption{Sky area of the \emph{IBIS/ISGRI\/} survey as a
  function of the limiting flux for the $4.5\sigma$-source
  detection.\label{fig:area}}
\end{figure}

The sky area covered by the \emph{IBIS/ISGRI\/} survey was
effectively restricted to the region for which achieved
sensitivity was better than 1 mCrab ($\simeq1.2\times 10^{-11}$
\flux) at the $4.5\sigma$ significance level in the 20--60 keV
energy band. The angular size of this region is of about $640$
deg$^2$. The survey area as a function of flux for sources with
$S/N>4.5$ is shown in Fig.~\ref{fig:area}.  Note
that the central part of the survey (with the area of $\sim240$
deg$^2$), where the LMC is located, has been covered with a
sensitivity better than $0.5$ mCrab in the same band.

%==================================================
%==================================================

\begin{figure}
\includegraphics[width=0.9\columnwidth,bb=55 270 550 710,clip]{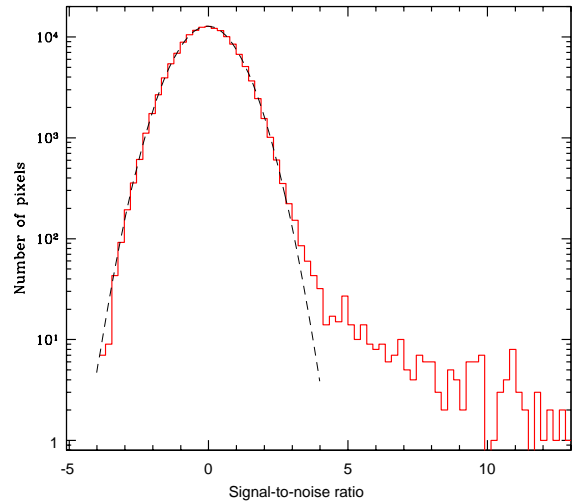}

\caption{Signal-to-noise ratio distribution of pixels in the
  20--60 keV image of the LMC (histogram). The dashed line
  represents the normal distribution with unit variance and zero
  mean.\label{fig:sign_distr}}
\end{figure}

One of the main purposes of this survey was searching for
previously unknown sources. The detection threshold was
estimated assuming Gaussian statistics for pixel values in the
accumulated (mosaic) sky image of the LMC. The total area of the
image is $\sim640$ deg$^2$ and, taking the \emph{IBIS\/} angular
resolution into account, we gathered $\sim1.6\times 10^{4}$
independent pixels. With this assumption, the formal detection
threshold at the signal-to-noise ratio $S/N=4$ allows at most
one false detection. However, the very deep \emph{IBIS/ISGRI}
mosaic images may be affected by a systematic noise or some
artifacts caused by the imperfect sky reconstruction
\citep{kri2010b}. In spite of the absence of strong apparent
systematic noise (see Fig.\,\ref{fig:sign_distr}) we set a more
conservative threshold $S/N=4.5$, that diminishes the
probability of the false detection by a factor more than 10.

To obtain an accurate localization of newly revealed hard X-ray
sources we used data from the follow-up observations performed
with the \emph{XRT} telescope aboard the \emph{Swift}
observatory \citep{gehr2004}.

\section{Survey}

%==================================================
%==================================================

%==================================================
\begin{figure*}
\includegraphics[width=0.9\textwidth,bb=35 145 575 645,clip]{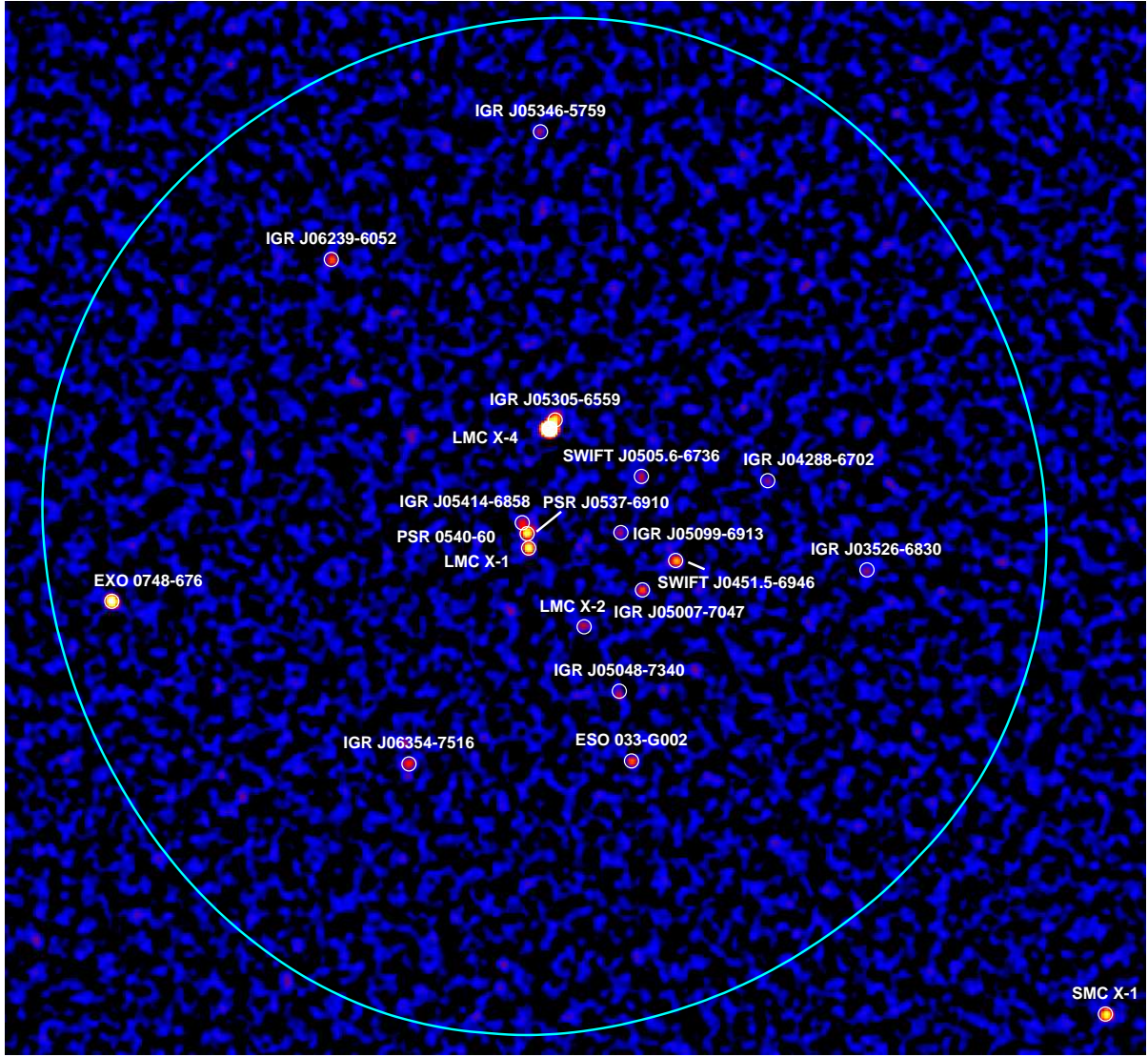}

\caption{Mosaic image ($S/N$ ratio map) of the LMC field obtained with
  \emph{IBIS/ISGRI} in the 20--60 keV energy band in
  2003-2012. The cyan contour restricts the area with the
  sensitivity better than 1 mCrab (for detection at
  $4.5\sigma$). All the sources with $S/N>4.5$ are
  labeled. \label{fig:map_large}}
\end{figure*}
%==================================================

%==================================================
\begin{table*}

   \caption{The catalogue of sources detected with $S/N>4.5$
     during the \emph{INTEGRAL} survey of the LMC field$^a$\label{tab:srclist}}
%   \medskip
   \hspace{-7mm}\begin{tabular}{l|c|r|r|l}
     \hline
     \hline
  Name                    &Significance$^b$& \multicolumn{2}{|c|}{Flux, mCrab} & Type$^c$, Other names  \\
                          &            & $20-60$ keV        & $3-20$ keV &        \\
      \hline
  LMC\,X-4                 &    266.1 & $21.04\pm0.08$ & $8.55\pm0.08$   &    HMXB  \\
  EXO\,0748-676            &     39.5 &  $6.88\pm0.17$ &                 &    LMXB, MW \\
  LMC\,X-1                 &     25.8 &  $1.98\pm0.08$ & $9.55\pm0.07$   &    HMXB, BH  \\
  IGR\,J05305-6559         &     23.2 &  $1.84\pm0.08$ & $0.69\pm0.08$   &    HMXB, EXO\,053109-6609 \\
  PSR\,0540-69             &     22.0 &  $1.68\pm0.08$ & $1.28\pm0.07$   &    PSR   \\
  IGR\,J06239-6052         &     11.1 &  $1.33\pm0.12$ &                 &    Sy2, z=0.0405, ESO\,121-IG 028 \\
  SWIFT\,J0451.5-6949      &     14.9 &  $1.22\pm0.08$ & $0.73\pm0.09$   &    HMXB  \\
  ESO\,033-G002            &     11.7 &  $1.14\pm0.10$ & $0.60\pm0.16$   &    Sy2, z=0.0184 \\
  IGR\,J06354-7516$^d$     &      8.7 &  $0.87\pm0.10$ &    &    PKS 0637-752, z=0.653, SWIFT\,J0635.9-7515 \\
  IGR\,J05346-5759         &      6.2 &  $0.87\pm0.14$ &    &    CV, MW  \\
  IGR\,J05007-7047         &     10.0 &  $0.81\pm0.08$ & $0.46\pm0.08$   &    HMXB  \\
  IGR\,J05414-6858         &      8.0 &  $0.61\pm0.08$ &    &    HMXB  \\
  IGR\,J03526-6830         &      4.8 &  $0.56\pm0.12$ &    &    BL Lac, PKS\,0352-686, IGR\,J03532-6829 \\
  SWIFT J0505.6-6736       &      6.4 &  $0.50\pm0.08$ &    &    Galaxy, 2MASX\,J05052442-6734358 \\
  LMC\,X-2                 &      6.1 &  $0.49\pm0.08$ & $16.90\pm0.09$   &    LMXB  \\
  IGR\,J04288-6702$^d$     &      5.3 &  $0.49\pm0.09$ &    &     \\
  IGR\,J05099-6913$^d$     &      6.1 &  $0.47\pm0.08$ &    &     \\
  IGR\,J05048-7340$^d$     &      5.3 &  $0.46\pm0.09$ &    &    Galaxy, z=0.0148, ESO\,033 -G011  \\
  PSR\,J0537-6910          &      5.8 &  $0.44\pm0.08$ & $0.25\pm0.06$    &    PSR  \\
  LMC\,X-3$^e$             &      2.4 &  $0.20\pm0.08$ & $11.38\pm0.11$   &    HMXB, BH  \\

     \hline

    \end{tabular}
    \medskip
    \begin{tabular}{ll}
    $^a$ & sources sorted by their flux in the 20--60 keV band
      and listed in descending order\\
    $^b$ & in the 20--60 keV energy band\\
    $^c$  & HMXB --- high-mass X-ray binary, LMXB --- low-mass
      X-ray binary, CV --- cataclysmic variable, PSR --- isolated Crab-like pulsar, \\
          & BH --- black hole, MW --- the Milky Way galaxy\\
    $^d$  & source discovered in this survey \\
    $^e$  & included due to significant detection in the
      standard (3--20 keV) X-ray band by the \emph{JEM-X} telescope\\
    \end{tabular}
\end{table*}
%=================================================

%==================================================
\begin{figure*}
\hbox{
\includegraphics[width=0.66\textwidth,bb=35 250 575 540,clip]{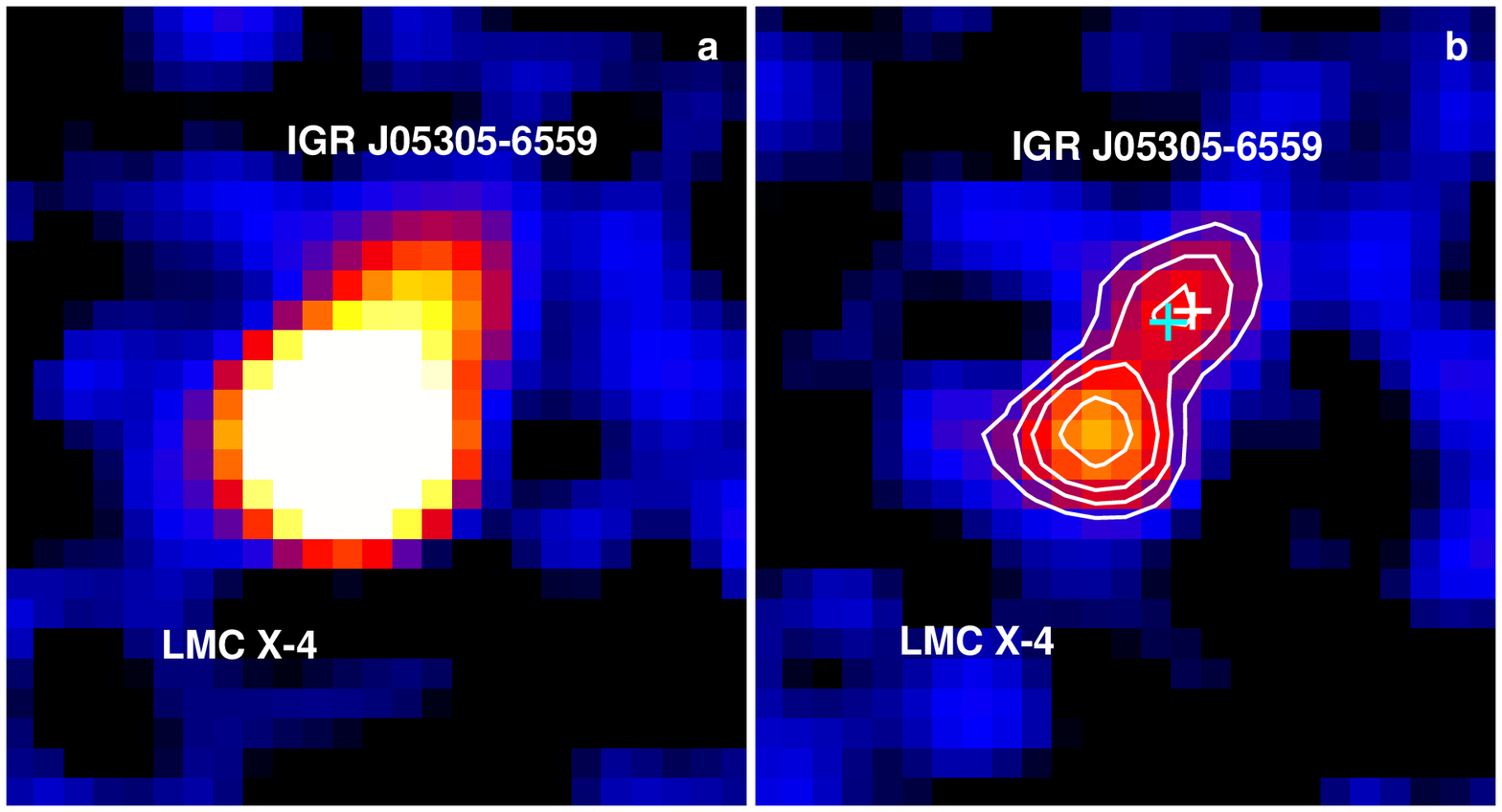}
\includegraphics[width=0.33\textwidth,bb=35 250 305 540,clip]{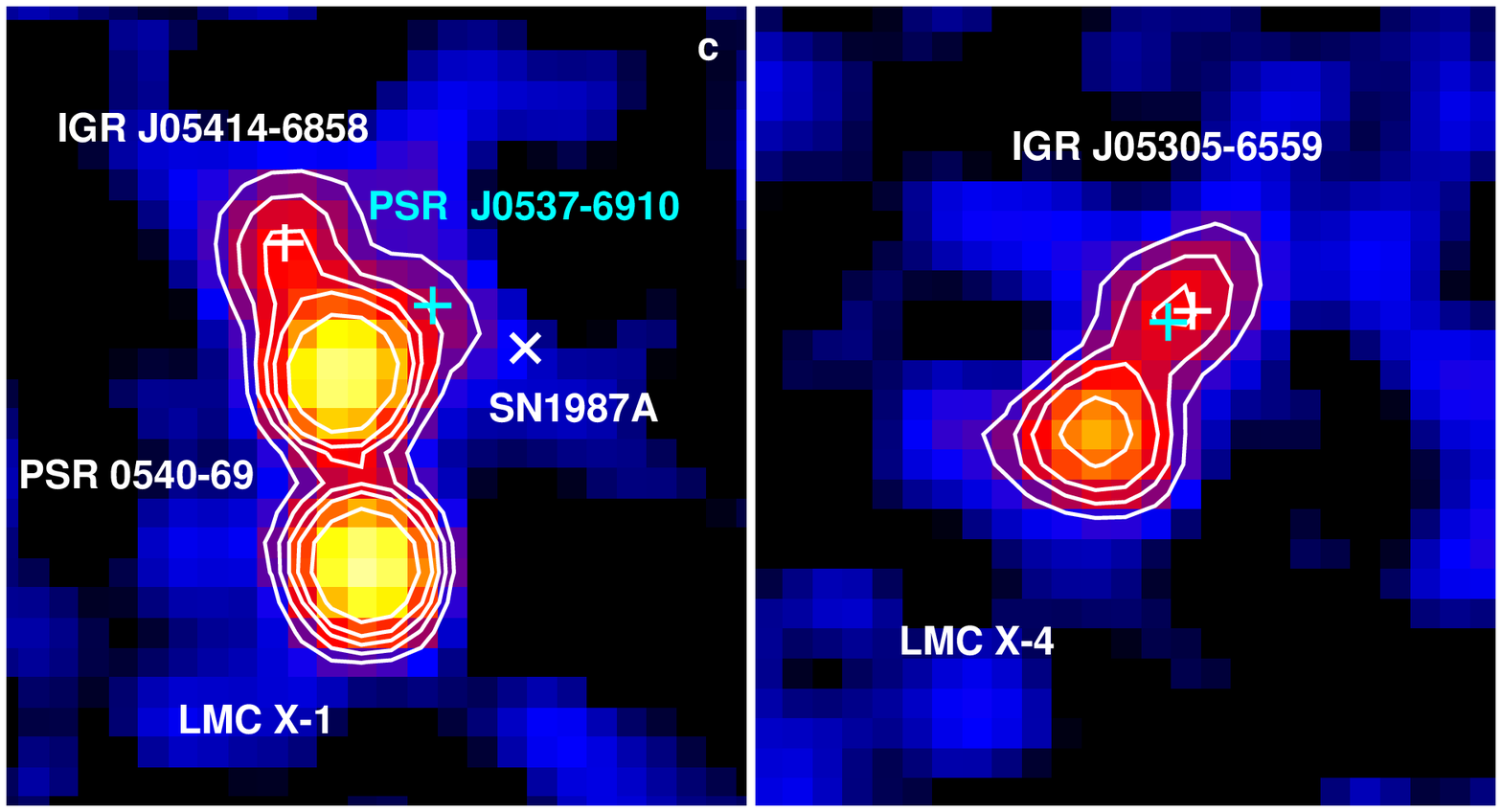}
}

\caption{Enlarged \emph{IBIS/ISGRI} maps of two crowded regions
  in the LMC: near the X-ray pulsar {LMC\,X-4} (a,b) and
  near the Crab-like pulsar {PSR\,0540-69} (c). The cyan
  and white crosses in the panel (b) indicate the positions of
  the soft X-ray sources {EXO\,053109-6609} and
  {XMMU\,J053041.1-660535}, respectively. Contours
  correspond to the $S/N$ ratio levels at $3.0, 4.7, 6.4$ and
  $9.8\sigma$. The cyan cross in the panel (c) indicates a
  position of the pulsar {PSR J0537-6910}, the white one
  --- a position of {IGR\,J05414-6858} discovered by us in
  2010. The position of the Supernova 1987A remnant is shown by
  a symbol X. Contours correspond to the $S/N$ ratio levels of
  $3.0, 4.7, 6.4, 8.1$ and $11.5\sigma$. See text for
  details. \label{fig:map_zoom}}
\end{figure*}
%==================================================

The \emph{IBIS/ISGRI} mosaic image of the LMC region in the
20--60 keV X-rays is shown in Fig.\,\ref{fig:map_large}. The
sources detected at the signal-to-noise ratio $S/N\ga4.5$ are
listed in Table\,1 (LMC\,X-3 is included in this table due to its
confident detection with the \emph{JEM-X} telescope). Because of the
large field of view of the \emph{IBIS} telescope the size of the
sky region covered with the sensitivity better than 1 mCrab exceeds
significantly that of the LMC itself ($\sim640$
deg$^2$ vs $\sim100$ deg$^2$, see Fig.\,\ref{fig:map_large},
cyan contour). Therefore there are a number of sources not related
to the LMC among those significantly detected in
Fig.\,\ref{fig:map_large} and listed in
Table\,\ref{tab:srclist}.

Below we discuss all the detected sources and their identifications.

\subsection*{Newly discovered sources}

Four new hard X-ray sources with $S/N>4.5$ were detected in the
mosaic image. These sources are labeled in Table\,\ref{tab:srclist} by
a letter $^{d}$. Their coordinates were determined with
\emph{IBIS/ISGRI} as follows (epoch J2000, uncertainty of
$\sim4$\arcmin):

\medskip
\begin{tabular}{ll@{}l@{}l l@{}l@{}l@{}}
    Name  &\multicolumn{3}{c}{R.A.} &\multicolumn{3}{c}{Dec.} \\
  IGR\,J04288-6702    &   04$^{\rm h}$& 28$^{\rm m}$&48\fs2&  -67\deg &02\arcmin&24\arcsec\\
  IGR\,J05048-7340    &   05 &04&46.6 & -73& 40 &01 \\
  IGR\,J05099-6913    &   05 &09&56.6 & -69& 13 &16 \\
  IGR\,J06354-7516    &   06 &35&25.9 & -75& 16 &55  \\

\end{tabular}
\medskip

To identify these sources and establish their nature we made an
analysis of available data from X-ray observatories
\emph{ROSAT}, \emph{XMM-Newton} and \emph{Swift/XRT} and searched
in different databases (\emph{SIMBAD, NED}, etc.). Two of the new sources were identified with known extragalactic objects:
IGR\,J06354-7516 --- with the quasi-stellar object PKS\,0637-752
located at redshift z=0.653, and IGR\,J05048-7340 --- with
the galaxy ESO\,033-G011 at redshift z=0.0148.

The sky region around IGR\,J05099-6913 was observed many times
with the \emph{XRT} telescope of the \emph{Swift} observatory
(ObsID. 00030348, 00037825, 00038031, 00090245 for a target
''LMC Nova 2005'', a total exposure is of $\sim48$ ks). No
source was detected at the significant level during these
observations inside the \emph{IBIS} error circle of
IGR\,J05099-6913, including the Nova position. The nearest soft
X-ray source has coordinates R.A.$=05^h$09$^m$15\fs6,
Dec.$=-69$\deg08\arcmin07\arcsec (J2000, uncertainty
$\simeq4$\arcsec) and located $\sim6$\arcmin\ away from the
\emph{IBIS/ISGRI} position of IGR\,J05099-6913. It is unlikely,
that this object is a soft X-ray counterpart of IGR\,J05099-6913
because of: 1) formally, the separation of $\sim6$\arcmin
exceeds the uncertainty of the \emph{IBIS/ISGRI} coordinates, 2)
the \emph{Swift/XRT} source is very soft (in particular, it is
practically undetectable in the 6--10 keV energy band).

The source IGR\,J04288-6702 was observed by \emph{Swift/XRT}
with a total exposure of $\sim10$ ks in October 2010
(ObsId. 00040907). Two faint soft X-ray sources with coordinates
R.A.$=04^h$29$^m$47\fs2, Dec.$=-67$\deg03\arcmin21\arcsec\ and
R.A.$=04^h$27$^m$49\fs1,
Dec.$=-67$\deg04\arcmin36\arcsec\ (J2000, uncertainties
4--5\arcsec) were detected at distances of $\sim6$\arcmin\ from
the \emph{IBIS/ISGRI} position. Based on these data and taking
into account a large separation between sources positions it is
difficult to make any conclusion about the nature of
IGR\,J04288-6702. We can mention only that the intensity of the
former \emph{XRT} source is $\sim5$ times larger than that of
the latter one and both of them are confidently detected in the 6--10 keV band.

\subsection*{High-mass X-ray binaries}
%==================================================
\begin{table*}
   \caption{Fluxes in the 20--60 keV energy band from the known
     hard X-ray sources in the LMC field not included in Table\,
     \ref{tab:srclist} but still appearing in the
     \emph{IBIS/ISGRI} mosaic image with
     $S/N\geq3\sigma$ \label{tab:fluxes}}

   \hspace{0mm}\begin{tabular}{l|c|r|l}
     \hline
     \hline
  Name                    &Significance$^a$& Flux$^a$, mCrab & Type, Other names  \\
      \hline

%  ESO 157-G023            &      3.7 &  $1.02\pm0.28$ &    Sy2, z=0.0435   \\
  1H 0419-577             &      3.4 &  $0.76\pm0.22$ &    Sy1, z=0.104 \\
  SWIFT J0450.7-5813      &      3.8 &  $0.58\pm0.15$ &    Sy1.5, z=0.0907, RBS 594\\
  ABELL 3266              &      4.1 &  $0.52\pm0.13$ &  Cluster of galaxies\\
  SWIFT J0747.9-7327      &      3.2 &  $0.49\pm0.15$ &    Galaxy, z=0.0367 \\
  SWIFT J0634.7-7445      &      3.9 &  $0.39\pm0.10$ &    1RXS\,J063401.1-744629 \\
  IGR\,J06233-6436         &      4.0 &  $0.37\pm0.09$ &    Sy1, z=0.1289, PMN J0623-6436 \\
%  1RXS J061148.5-662430   &      4.1 &  $0.34\pm0.08$ &    ??? \\
  SWIFT J0609.5-6245      &      3.0 &  $0.30\pm0.10$ &    Galaxy \\

     \hline
    \end{tabular}
    \medskip

    \begin{tabular}{ll}
    $^a$  & in the 20--60 keV energy band \\
    \end{tabular}

\end{table*}
%==================================================

From 19 hard X-ray sources listed in Table \ref{tab:srclist} six
ones are high-mass X-ray binaries (HMXBs), located in the
LMC. Four of them (LMC\,X-4, LMC\,X-1, IGR\,J05007-7047 and
IGR\,J05009-7044) were previously reported by
\citet{kri2010a,bird2010}. The type of SWIFT\,J0451.5-6949 was
established after the discovery of X-ray pulsations with a
period of $\simeq187$ s in the source light curve and its
identification with the blue star located in the LMC
\citep{beard2009}. The source IGR\,J05414-6858 was discovered in
the course of this survey in June 2010 and reported earlier
\citep{greb2010}. The follow-up observations with the
\emph{Swift/XRT} telescope allowed to improve the source
localization and proposed its optical counterpart
\citep{lut2010} while the observations with the
\emph{Swift/UVOT} and 2.2-m \emph{MPG/ESO} telescopes identified it
with a Be X-ray binary \citep{rau2010}. Using data of the
\emph{XMM-Newton} observations of this source carried out during
its outburst in 2011, \citet{stu2012} revealed X-ray pulsations
with a period of $\sim4.4$ s in its light curve and classified
the source as a X-ray pulsar in a high-mass X-ray binary system. The
sixth HMXB, IGR\,J05305-6559, shows a much more complex
identification process.  It was previously reported by
\citet{kri2010a} in the \emph{IBIS/ISGRI} 7-year all-sky survey with
the remark ''confusion'' due to its flux contamination by the
nearby bright X-ray pulsar LMC\,X-4. This made an accurate
determination of the IGR\,J05305-6559 position and its
identification to be complex (e.g.,
Fig.\,\ref{fig:map_zoom}a). Fortunately, the X-ray pulsar LMC\,X-4
shows a superorbital variability with a period of $\sim30.5$
days and a rather long interval of switching-off (the
''off''-state). We accumulated a mosaic image of this region
using the \emph{IBIS/ISGRI} data related only to this
''off''-state that allowed to separate IGR\,J05305-6559 from LMC\,X-4
(Fig.\ref{fig:map_zoom}b).  It is obvious from the figure
that the position of IGR\,J05305-6559 is consistent with the
positions of two soft X-ray sources --- EXO\,053109-6609 (shown
by a cyan cross) and XMMU\,J053041.1-660535 (a white
cross). Taking into account that the luminosity of
EXO\,053109-6609 in the 2--10 keV band is $\sim20$ times higher
than the luminosity of XMMU\,J053041.1-660535 \citep{shtyk2005}
we concluded that EXO\,053109-6609 is likely to be a soft X-ray
counterpart of IGR\,J05305-6559 which is thus a high-mass X-ray
binary system.

It is important to note, that all the discussed hard X-ray HMXBs
have a luminosity higher than $2\times10^{36}$ erg s$^{-1}$ in
the 20--60 keV energy band.  Taking this into account the
apparent small number of HMXBs registered in the LMC occurs
indeed to be consistent with the number of HMXBs of the same
brightness registered in our Galaxy. This follows from the
similarity of the luminosity functions of HMXBs in these two
galaxies, the LMC and the Milky Way (Lutovinov et al., in
preparation).

\subsection*{Extragalactic objects}

The next large group of sources detected by \emph{IBIS/ISGRI}
includes active galactic nuclei and other extragalactic
sources. In total six such objects were significantly detected
in this survey. Two of these sources (IGR\,J04288-6702 and
IGR\,J05048-7340) are newly discovered and discussed above. Two
more sources are active galactic nuclei of Seyfert 2 type
(ESO\,033-G002 and IGR\,J06239-6052) located at redshifts
$z\simeq0.018$ and $\simeq0.04$, respectively (see
Table\,\ref{tab:srclist}). The latter two are a blazar
(IGR\,J03526-6830) and a galaxy (SWIFT\,J0505.6-6736). It is
worth to note that the total number of extragalactic objects
detected in the direction to the LMC is smaller compared to
other directions. This could be an indication of the global
non-uniformity of the matter distribution in the local Universe
\citep[see][for details]{lut2012}.

\subsection*{Other types of detected sources}

%==================================================
\begin{figure*}
\includegraphics[width=0.9\textwidth,bb=35 145 575 645,clip]{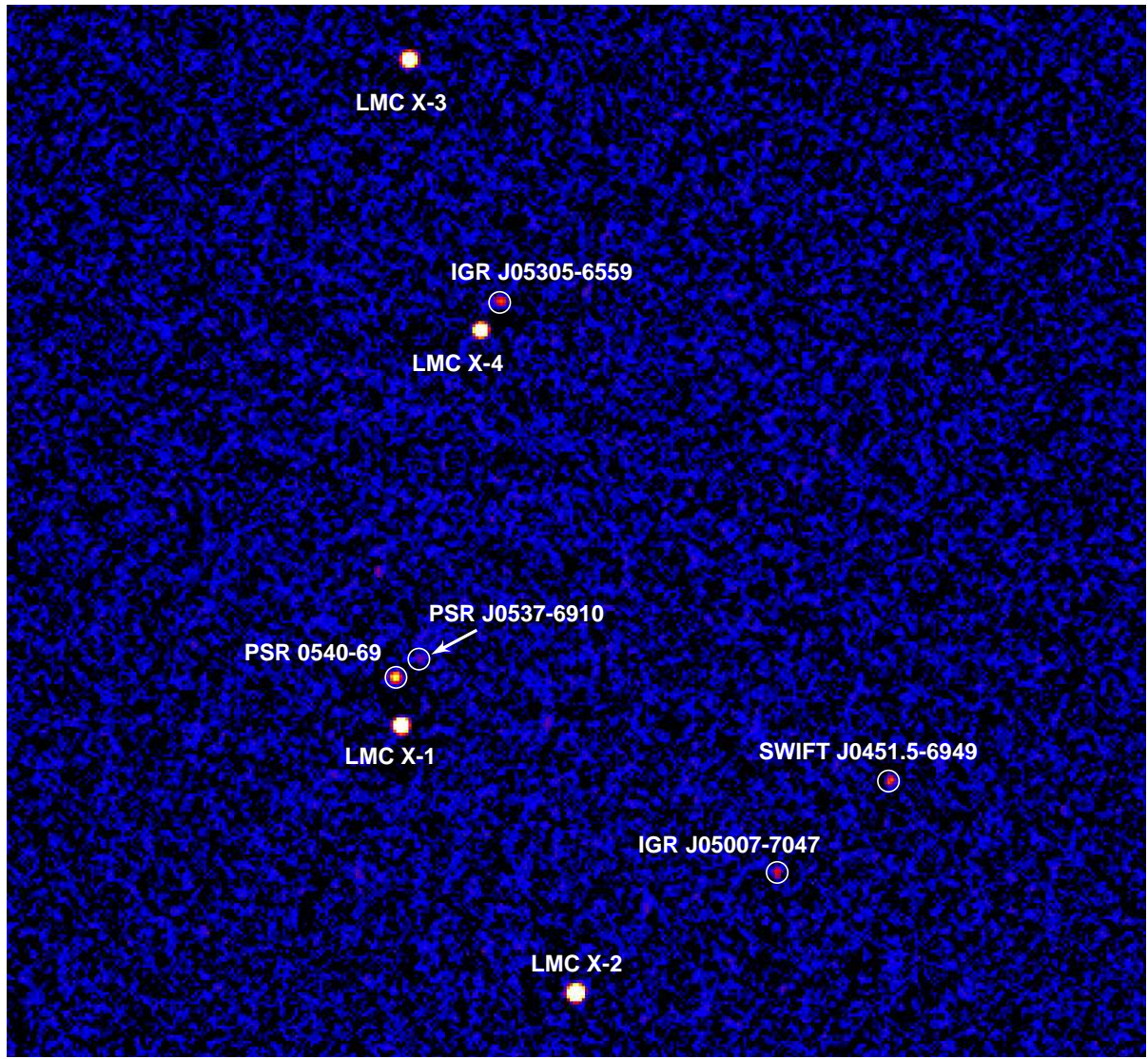}

\caption{Mosaic image ($S/N$ ratio map) of the LMC field
  obtained by the \emph{INTEGRAL/JEM-X} telescope in the 3--20
  keV energy band. All significantly detected sources are
  labeled. \label{fig:map_jemx}}
\end{figure*}
%==================================================

Although HMXBs is the main population of X-ray sources in the
LMC \citep[see, e.g.,][]{shtyk2005}, \emph{INTEGRAL} has
detected one low-mass X-ray binary (LMC\,X-2) and two rotating
powered Crab-like pulsars --- PSR\,0540-69 and
PSR\,J0537-6910. The former pulsar is the well known source
previously reported by other missions and included in the
\emph{INTEGRAL} and \emph{Swift} all-sky catalogues. A
detailed study of the region around this source showed that the
contours of its $S/N$ ratio is not nearly circular, as
usual, but instead is oblong in two directions
(Fig.\,\ref{fig:map_zoom}c). In the one direction the oblongness
is associated with the high mass X-ray binary IGR\,J05414-6858
discovered by \citet{greb2010}. In the second direction the oblongness
coincides with the position of another Crab-like pulsar
PSR\,J0537-6910. The ratio of the fluxes measured by
\emph{IBIS/ISGRI} from these pulsars ($\sim4$) coincides with
the ratio of their fluxes measured by \emph{XMM-Newton} in the
softer X-ray band \citep{shtyk2005}, that is the result of the
power law spectra of both sources with similar photon
indexes. This confirms our conclusion that PSR\,J0537-6910 is
most likely responsible for the detected oblongness of the $S/N$
ratio contours. Note, that this source was detected in hard
X-rays for the first time.

Finally, two X-ray sources owned to our Galaxy, EXO\,0748-676
(LMXB) and IGR\,J05346-5759 (CV), and one source owned to the
Small Magellanic Cloud, \mbox{SMC\,X-1} (HMXB at flux
$=28.1\pm1.3$ mCrab), located outside the 1 mCrab region, were
significantly detected with \emph{IBIS/ISGRI} in the survey.

Fig.\,\ref{fig:map_jemx} shows the LMC mosaic image obtained with the
\emph{JEM-X} telescope in the 3--20 keV energy band. Due to the smaller
field of view of \emph{JEM-X} this image covers only the central part
of the hard X-ray survey which corresponds roughly to the LMC galaxy itself.
Eight sources from the \emph{IBIS/ISGRI} catalogue were significantly
detected by \emph{JEM-X} (see Table\,\ref{tab:srclist}) and
in addition --- the very bright black-hole binary LMC\,X-3 not
detected by \emph{IBIS}. The Crab-like pulsar PSR\,J0537-6910 was
marginally detected and its flux can be considered as an upper limit.

Besides the above discussed sources confidently ($S/N>4.5$)
detected with \emph{IBIS/ISGRI}, a number of other ones were
observed in previous hard X-ray surveys within the considered region
of the \emph{IBIS/ISGRI} 1-mCrab sensitivity. For completeness
of our survey, for the sources registered with the $S/N>3.0$ we
estimated fluxes at their positions in the 20--60 keV band (see Table\,\ref{tab:fluxes}). In total, the
fluxes from seven such sources are presented. Six of them are
extragalactic objects, the origin of yet another source is still
unclear.

\section{Broad-band X-ray spectra}

%==================================================
\begin{figure*}
\includegraphics[width=0.9\textwidth,bb=20 160 565 690,clip]{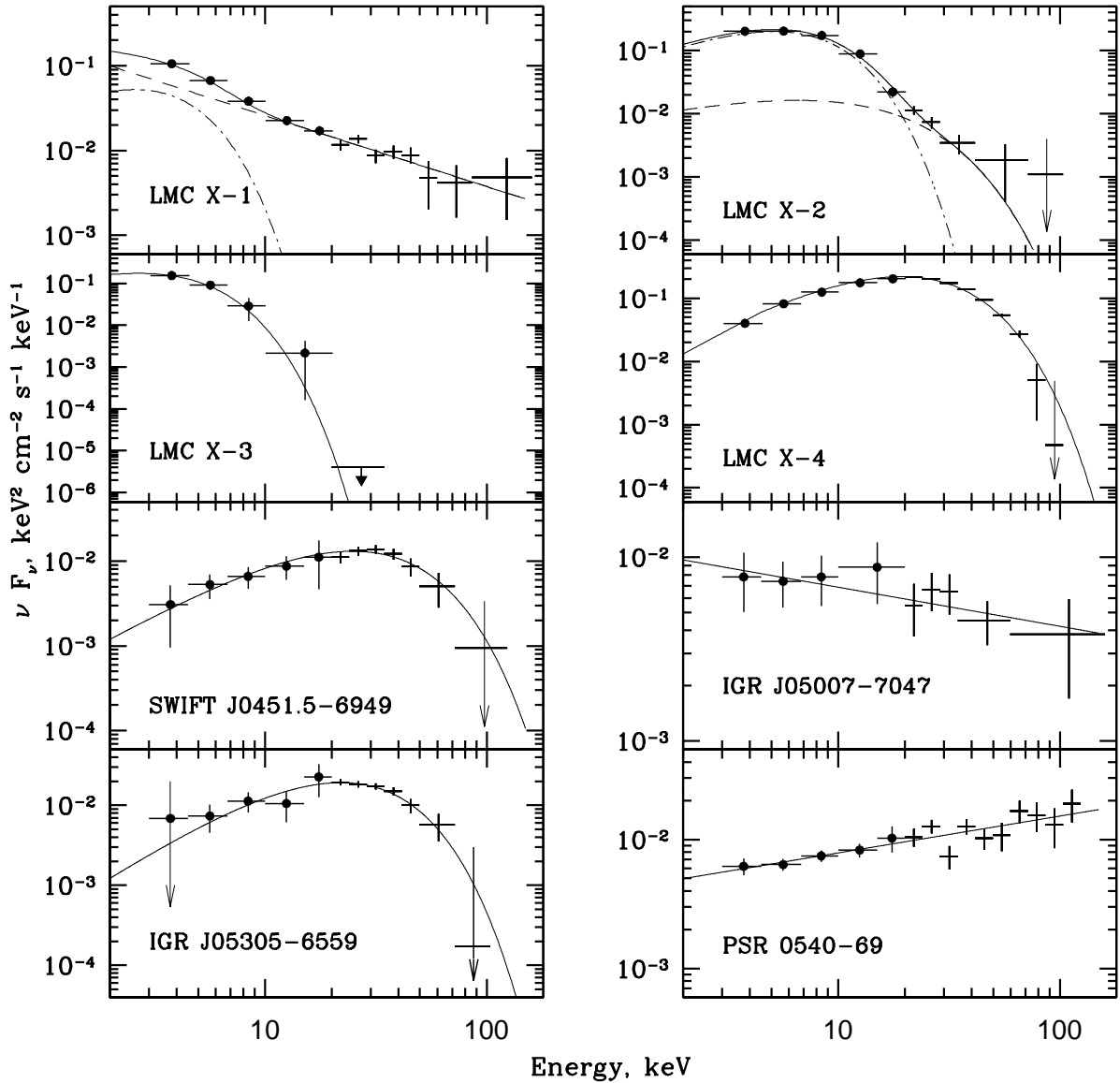}

\caption{Energy spectra of 8 X-ray sources confidently detected
  in the LMC both with \emph{JEM-X} and \emph{IBIS/ISGRI}
  instruments. Filled circles show results of the \emph{JEM-X}
  measurements, crosses --- the \emph{IBIS/ISGRI} ones. The
  best-fit models are shown by solid lines. Dashed and
  dashed-dotted lines represent different components of complex
  spectral models (see text for details).  The bold arrow shows
  a $3\sigma$ upper limit for the hard X-ray emission from LMC\,X-3.\label{fig:spectra}}
\end{figure*}
%==================================================

The broad-band spectra in the 3--100 keV energy band of 8 X-ray
sources located in the LMC and confidently detected with both
the \emph{JEM-X} and \emph{IBIS/ISGRI} instruments are presented
in Fig.\,\ref{fig:spectra}. Results of \emph{JEM-X} and
\emph{IBIS} measurements are shown by filled circles and
crosses, respectively; corresponding best-fit models are shown
by solid lines. The model parameters are given in
Table\,\ref{tab:models}.

It follows from Table\,\ref{tab:srclist} that two high-mass
X-ray binaries harboring black holes were detected in the
survey, namely LMC\,X-1 and LMC\,X-3. The spectrum of LMC\,X-1
can be successfully described by a two-component model
consisting of a disk black-body and a power law (their
contribution are shown in Fig.\,\ref{fig:spectra} by dash-dotted
and dashed lines, respectively). Such a model as well as its
parameters in Table\,\ref{tab:models} are typical both for
objects of this type and LMC\,X-1 itself
\citep{tanaka96,rem2006,ruh2011}. As it was mentioned above
another HMXB/BH system, LMC\,X-3, was significantly detected
only by the \emph{JEM-X} telescope (see
Table\,\ref{tab:srclist}). This is connected with the ultrasoft
state of this source, whose spectrum can be approximated by a
one-component disk black-body model with the temperature
$kT_{in}\simeq1.1$ keV (Fig.\,\ref{fig:spectra}). The spectrum
of LMC\,X-2 which is a low-mass X-ray binary with a neutron star is also typical for
objects of this type. Formally it can be approximated with two
components: a disk black-body one with the temperature
$kT_{in}\simeq2$ keV and a bremsstrahlung of an optically thin plasma with the
temperature $kT_{br}\simeq10$ keV, that reflects a contribution of different
emission regions (boundary layer, accretion disk, hot corona, etc.) to the total spectrum
\citep{gs2002, gilf2003, sul2006}. The pulsar PSR\,0540-69 spectrum can be modeled with
a power law with the photon index $\Gamma\simeq1.7$, that is
typical for these objects in this energy range. The remaining
four sources are high-mass X-ray binaries with neutron
stars. The spectra of three of them (LMC\,X-4,
SWIFT\,J0451.5-6949, IGR\,J05305-6559) are very similar each
other and can be approximated by a power law with a high energy
cut-off as reported in Table\,\ref{tab:models}. The fourth
source, IGR\,J05007-7047, is fainter, therefore its spectrum was
obtained with large uncertainties in most of the energy channels
(Fig.\,\ref{fig:spectra}). It doesn't require complex spectral
modeling and can be described by a simple power law. No
cyclotron absorption lines were detected in the spectrum of the
known X-ray pulsar LMC\,X-4 that is consistent with the previous
reports \citep{tsy2005}.
%==================================================

\begin{table*}

\caption{Parameters of best-fit approximations of spectra for 8 X-ray sources shown in
  Fig.\,\ref{fig:spectra} \label{tab:models}}

\begin{tabular}{lcccc}
\hline\hline
    Name  & $kT_{in}^a$  & $kT_{br}^b$& $\Gamma^c$ & $E_{c}^d$ \\
                & keV  &  keV & &  keV \\
\hline
  LMC\,X-1            &  $1.08\pm0.06$ &  & $2.8\pm0.2$ & \\
  LMC\,X-2            &  $2.17\pm0.03$  & $10\pm1$   &  & \\
  LMC\,X-3            &  $1.14\pm0.04$  &    &  & \\
  LMC\,X-4            &    &   &$0.10\pm0.05$ & $10.4\pm0.6$ \\
  SWIFT\,J0451.5-6949 &    & &  $0.5\pm0.5$ & $16.0\pm5.0$ \\
  IGR\,J05007-7047    &    &   &$2.2\pm0.3$ & \\
  IGR\,J05305-6559    &    &   & $0.15\pm0.65$  &  $12\pm4$ \\
  PSR\,0540-69        &    &   & $1.7\pm0.1$  & \\
\hline
\end{tabular}    \medskip

    \begin{tabular}{ll}
    $^a$  & inner temperature in the disc black-body model\\
    $^b$  & bremsstrahlung  temperature\\
    $^c$  & photon index of the power-law approximation\\
    $^d$  & energy of the high energy cut-off in the spectrum\\
    \end{tabular}
\end{table*}
%==================================================

Finally note that the broad-band spectra of SWIFT\,J0451.5-6949,
IGR\,J05007-7047 and IGR\,J05305-6559, are obtained and
presented here for the first time.

\section{Summary}

We present results of the deep hard X-ray survey of the LMC field carried out with the \emph{INTEGRAL}
observatory in the period 2003--2012 for a total exposure of $\sim7$ Ms.
The main results can be summarized as follows:

\begin{itemize}
\item During this survey 20 X-ray sources were detected by
  \emph{IBIS/ISGRI} in the 20--60 keV band at the significance
  level $S/N>4.5$ with one more source being detected by the
  \emph{JEM-X} telescope only in the 3--20 keV band. In total
  \emph{JEM-X} detected 10 sources.

\item Ten of these 21 sources belong to the LMC (7 HMXBs, 2
  PSRs, 1 LMXB), six sources have an extragalactic origin, two
  belong to our Galaxy, and one more --- to the Small
  Magellanic Cloud. Two sources are still unidentified.

\item Four new hard X-ray sources were discovered in this survey
  and reported here for the first time: two of them are
  extragalactic objects, the nature of other two sources is
  still needed to be established. One more source,
  IGR\,J05414-6858, discovered during this survey has been
  reported earlier.

\item We report for the first time the detection of hard X-rays
  from the Crab-like pulsar PSR\,J0537-6910.

\item The source IGR\,J05305-6559 detected by \emph{INTEGRAL} in
  the close vicinity of the very bright X-ray pulsar LMC\,X-4 is
  most likely to be identified with the high-mass X-ray binary
  EXO\,053109-6609.

\item Broadband spectra (3--100 keV) of 8 bright sources are
  presented and analyzed. For three of these sources such
  spectra are reproduced and presented for the first time. All three
  sources are belong to the class of HMXBs, two of them have a similar
  spectral shape -- power law with an exponential cutoff and one more --
  simple power law (probably owing to a lack of statistic).
\end{itemize}

\section*{Acknowledgments}

This work was supported by the Russian Academy of Sciences
(program ``Non-stationary phenomena in the Universe''),
the Russian Ministry of Science and Education (under State
contract 14.740.11.0611), the Russian President (grant
NSh-5603.2012.2), the Russian Foundation for Basic Research
(grants 10-02-01466 and 11-02-12285-ofi-m-2011),
and the Academy of Finland (grant 127512). The research was
based on the data obtained from the European (ISDC) and Russian Science
Data Centers for INTEGRAL. This work is mainly based on
observations of the INTEGRAL observatory, an ESA project with
the participation of Denmark, France, Germany, Italy,
Switzerland, Spain, the Czech Republic, Poland, Russia and the
United States.

\label{lastpage}

\end{document}